\documentclass[12pt]{iopart}
\usepackage{epsfig,color}
\newcommand{\gtrsim}{\,\rlap{\lower4pt\hbox{$\mathchar\sim$}}
\raise1.8pt\hbox{$>$}\,}
\newcommand{\lesssim}{\,\rlap{\lower4pt\hbox{$\mathchar\sim$}}
\raise1.8pt\hbox{$<$}\,}

\begin{document}
						
\title[Supernova $\nu$'s and $\overline\nu$'s: ternary luminosity diagram and spectral split patterns]
                {Supernova neutrinos and antineutrinos: ternary luminosity diagram
           and spectral split patterns}
   
\author{Gianluigi Fogli$^{1,2}$, 
		Eligio Lisi$^2$, 
		Antonio Marrone$^{1,2}$ and 
 		\\ Irene Tamborra$^{1,2}$ }
\address{$^1$~Dipartimento Interateneo di Fisica ``Michelangelo Merlin,'' \\
		Via Amendola 173, 70126 Bari, Italy} 
%%%%%%% ++++++++++
\address{$^2$~Istituto Nazionale di Fisica Nucleare, Sezione di Bari,\\
         Via Orabona 4, 70126 Bari, Italy}
%%%%%%% ++++++++++

%\date{October 2008}

\begin{abstract}
In core-collapse
supernovae, the $\nu_e$ and $\overline\nu_e$ species may experience collective flavor swaps 
to non-electron species $\nu_x$, within energy 
intervals limited by relatively sharp boundaries (``splits''). These phenomena
appear to depend sensitively upon the initial energy spectra and luminosities. 
We investigate the effect of generic variations of the fractional luminosities
$(l_e,\, l_{\overline e}, \, l_x)$ with respect to the usual ``energy equipartition'' case 
 $(1/6,\,1/6,\,1/6)$, within an early-time supernova scenario with fixed thermal 
spectra and total luminosity. We represent the constraint $l_e+l_{\overline e}+4l_x=1$ 
in a ternary diagram, which is explored via numerical experiments (in single-angle approximation)
over an evenly-spaced grid of points.  
In inverted hierarchy, single  splits arise in most
cases, but an abrupt transition to double splits is observed for a few points
surrounding the equipartition one. 
In normal hierarchy, collective effects turn out to be unobservable 
at all grid points but one, where single splits occur. 
Admissible deviations from equipartition may thus induce dramatic changes in
the shape of supernova (anti)neutrino spectra. The observed patterns 
are interpreted in terms of initial flavor polarization vectors
(defining boundaries for the single/double split transitions),
lepton number conservation, and minimization of potential energy.

\end{abstract}


\maketitle

%%%%%%%%%%%%%%%%%%%%%%%%%%%%%%%%%%%%%%%%%%%%%%%%%%%%%%%%%%%%%%%%%%%%%%%%%%%%%%%%%%%%%%%%%%%%%%%%%%%%%
\section{Introduction}
%%%%%%%%%%%%%%%%%%%%%%%%%%%%%%%%%%%%%%%%%%%%%%%%%%%%%%%%%%%%%%%%%%%%%%%%%%%%%%%%%%%%%%%%%%%%%%%%%%%%%

Supernova neutrinos and antineutrinos (SN $\nu$ and $\overline\nu$)
have long been studied as probes of both stellar and particle properties \cite{Raff,Expl}. 
In particular, core-collapse supernovae represent a unique laboratory for studying high-density 
(anti)neutrino-(anti)neutrino interactions and their associated flavor evolution \cite{Pant}. Recent results
from large-scale calculations \cite{Coll,Semi} have sparked a renewed interest on this topic, which 
is now the focus of a growing literature, as reviewed in \cite{Di08,Duan}. 
An interesting outcome of all these studies has been the finding of neutrino flavor changes
with similar (``collective'') features over extended energy ranges. 

In most supernova scenarios, the flavor evolution can often be reduced to an effective two-family 
framework $(\nu_e,\,\nu_x)$ governed by mass-mixing parameters $(\pm \Delta m^2,\,\theta_{13})$
\cite{Di08,Duan}. In usual
notation, $\nu_x$ denotes 
either $\nu_\mu$ or $\nu_\tau$, or a linear combination of $\nu_{\mu}$ and $\nu_{\tau}$ (and similarly for antineutrinos). 
The upper and
lower sign of  $\Delta m^2$ refer to normal and inverted hierarchy,
respectively. In these scenarios, collective effects are triggered by any nonzero 
value of the neutrino mixing angle $\theta_{13}$ in inverted hierarchy (in strict analogy with the fall of a 
inverted pendulum \cite{Pend}) and end up with a complete exchange (``swap'') of 
flavor $\nu_e\leftrightarrow\nu_x$ \cite{Split,Swap} above a ``split'' energy $E_c$ 
dictated by $\nu_e-\overline\nu_e$ number conservation.

In such effective $2\nu$ scenarios, 
the paradigm of a single spectral split in the neutrino sector---supported by detailed constructions 
in adiabatic approximation \cite{Smirnov}---has been challenged by the observation of another, low-energy split
in the antineutrino sector \cite{Miri,Tamb,Three}. 
Indeed, the joint occurrence of single $\nu$ and $\overline\nu$ single 
splits has been recently recognized as a rather general feature in inverted hierarchy, 
and as a novel possibility in normal hierarchy \cite{Mult}. Moreover, cases with
double $\nu$ and $\overline\nu$ splits have been identified as well \cite{Mult}. 
Relaxing the assumption of energy equipartition has been instrumental in obtaining 
such multiple split
cases. 

In general, the total energy luminosity $L_\mathrm{tot}$ is distributed over six ($3\nu+3\overline\nu$) species,  
%..........................
\begin{equation}
L_\mathrm{tot}=L_e + L_{\overline e}+4L_{x}\ ,
\label{Ltot}
\end{equation}
%.........................
or, equivalently,
%..........................
\begin{equation}
1=l_e + l_{\overline e}+4l_{x}\ ,
\label{Lcon}
\end{equation}
%.........................
where we have introduced the fractional luminosities
%..........................
\begin{equation}
 l_\alpha=L_\alpha/L_\mathrm{tot}\ \ (\alpha=e,\,\overline{e},\,x)\ ,
\end{equation}
%..........................  
with $l_x \equiv l_{\overline x}$ (we shall often ignore any distinction between $\nu_x$ and $\overline\nu_x$).
The usual assumption of ``energy equipartition'' amounts to take $l_e=l_{\overline e}=l_x$, namely,
%..........................
\begin{equation}
\mathrm{equipartition}\ \Leftrightarrow\ (l_e, \,l_{\overline e},\, 4l_{x})=(1/6,\,1/6,\,4/6)\ .
\label{unit}
\end{equation}
%.........................

\newpage

%%%%%%%%%%%%%%%%%%%%%%%%%%% FIGURE 1 %%%%%%%%%%%%%%%%%%%%%%%%%%%%%%%%
\begin{figure}[t]
\centering
\vspace*{1mm}
\hspace*{23mm}
\epsfig{figure=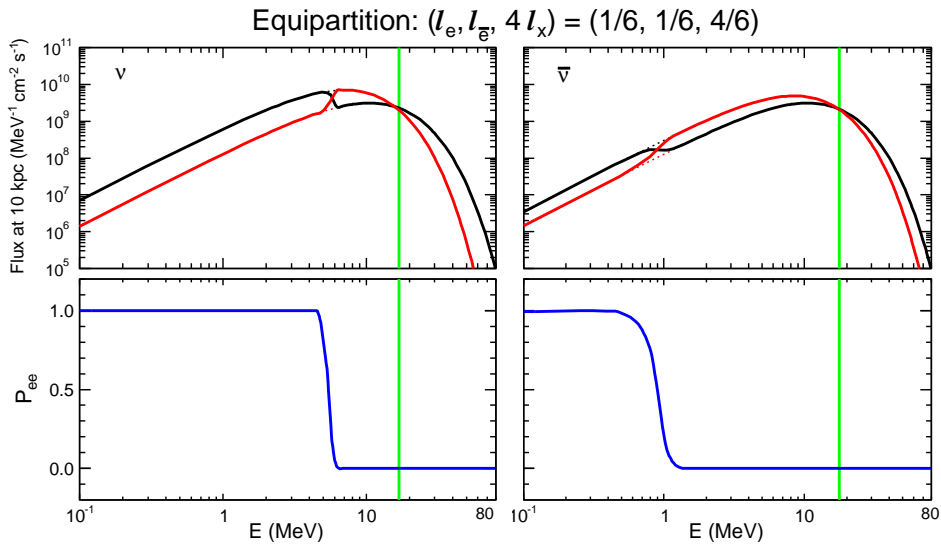,width=0.8\columnwidth}
\vspace*{-5mm}
\caption{Equipartition case  in inverted hierarchy. Left upper panel: 
flux spectra of $\nu_e$ (black, solid) and of $\nu_x$ (red, solid) at the end of collective effects;
dotted lines indicate initial spectra. All spectra are formally projected at a distance 
$d=10$~kpc. Left lower panel: electron flavor survival probability 
$P_{ee}$. Right upper and lower panel: as before, but for antineutrinos. Vertical green lines
mark the crossing energies where the $\nu_e$ (or $\overline\nu_e$) and $\nu_x$ fluxes 
are equal.
\label{fig01}}
\end{figure}
%%%%%%%%%%%%%%%%%%%%%%%%%%%%%%%%%%%%%%%%%%%%%%%%%%%%%%%%%%%%%%%%%%%%%%

Figure~\ref{fig01} shows typical results at the end of collective effects
(for inverted hierarchy), 
in a SN scenario endowed with the equipartition hypothesis, as also adopted in our previous papers \cite{Miri,Tamb,Three}. 
The $\nu_e$ and $\nu_x$ differential fluxes  
are swapped above a split energy $E_c\simeq 6$~MeV. A similar
swap occurs between $\overline\nu_e$ and $\overline\nu_x$, but at a lower (and not as sharply
defined) split energy $\overline E_c \sim 1$~MeV. No swap occurs in normal hierarchy 
for either $\nu$ or $\overline\nu$ (not shown).
Details of the SN scenario will be given in Sec.~\ref{Sec3}; here we just emphasize that 
significant (qualitative and quantitative) departures 
from such results can be induced by admissible
deviations from equipartition, such as those considered in \cite{Mult}. 
Therefore, it is worthwhile to investigate the spectral dependence upon the
initial SN luminosities.

In this work, we perform an extensive phenomenological investigation in
a reference SN scenario where all parameters are fixed, except for
the fractional luminosities $l_\alpha$. As described in Sec.~\ref{Sec2}, the constraint 
$l_e + l_{\overline e}+4l_{x}=1$    [Eq.~(\ref{Lcon})] 
is embedded in a ``ternary luminosity diagram,'' sampled through an evenly-spaced grid of points. 
Details of the SN model used are given in Sec.~\ref{Sec3}. Numerical results for inverted
hierarchy are described at length in Sec.~\ref{Sec4}. In particular, abrupt transitions
are observed from single to double split cases, across some regions of the
ternary diagram surrounding the equipartition case. An interpretation of the results
in terms of initial polarization vectors is offered in Sec.~\ref{Sec5}. 
Results 
for normal hierarchy are discussed in Sec.~\ref{Sec6}. Comments on
``more adiabatic'' scenarios are provided in Sec.~\ref{Sec7}. All our findings are briefly summarized
in Sec.~\ref{Sec8}, together with prospects for further work.

%%%%%%%%%%%%%%%%%%%%%%%%%%%%%%%%%%%%%%%%%%%%%%%%%%%%%%%%%%%%%%%%%%%%%%%%%%%%%%%%%%%%%%%%%%%%%%%%%%%
\section{Ternary Luminosity Diagram}
\label{Sec2}
%%%%%%%%%%%%%%%%%%%%%%%%%%%%%%%%%%%%%%%%%%%%%%%%%%%%%%%%%%%%%%%%%%%%%%%%%%%%%%%%%%%%%%%%%%%%%%%%%%%%

%%%%%%%%%%%%%%%%%%%%%%%%%%% FIGURE 2 %%%%%%%%%%%%%%%%%%%%%%%%%%%%%%%%
\begin{figure}[t]
\centering
\vspace*{1mm}
\hspace*{23mm}
\epsfig{figure=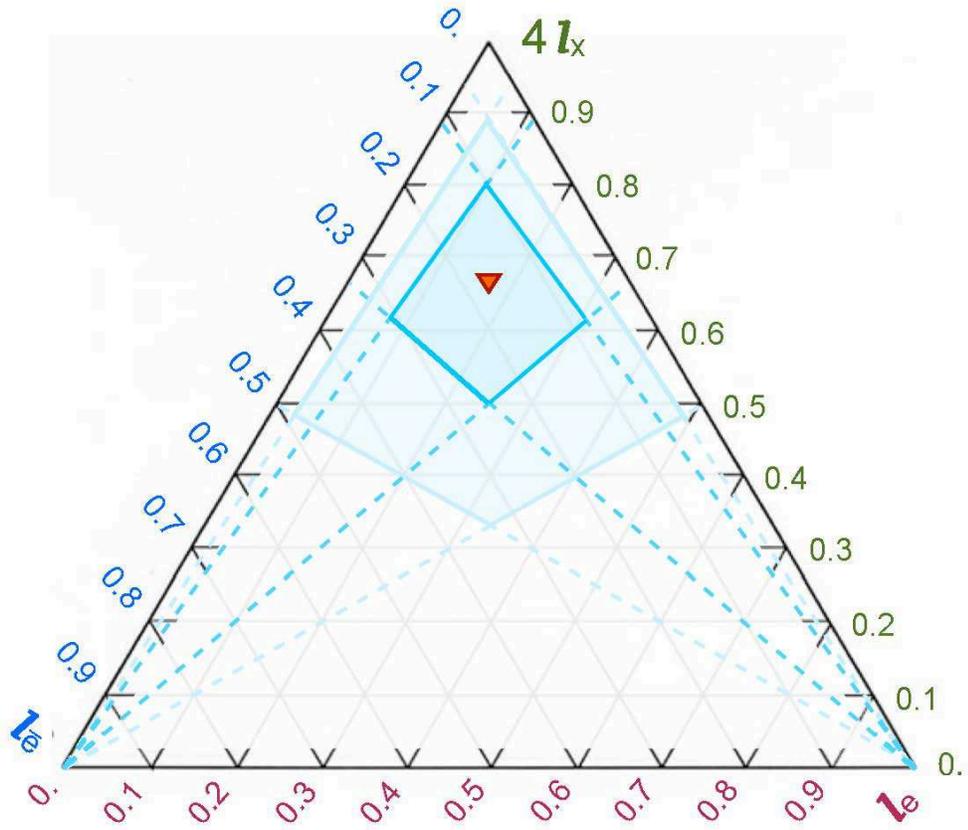,width=0.84\columnwidth}
\vspace*{-5mm}
\caption{Ternary luminosity diagram. Each point corresponds to fractional luminosities 
$(l_e,\,l_{\overline e},\,4l_x)$ subject to the constraint $l_e+l_{\overline e}+4l_x=1$.
The equipartition point is marked by a red triangle. The diagram is charted by an evenly-spaced grid of points. 
The small (large) shaded diamond correspond to ratios
$l_e/l_x$ and $l_{\overline e}/l_x$
differing from~1 by less than a factor of two (four).
\label{fig02}}
\end{figure}
%%%%%%%%%%%%%%%%%%%%%%%%%%%%%%%%%%%%%%%%%%%%%%%%%%%%%%%%%%%%%%%%%%%%%%

Unitarity constraints of the form $h_1+h_2+h_3=1$ can be conveniently represented in ternary diagrams 
by means of Viviani's theorem \cite{Viviani}, where the $h_i$'s represent the heights projected
by any point inside an equilateral triangle of total unit height. In our case [Eq.~(\ref{Lcon})], the heights 
are identified with the fractional neutrino energy luminosities: $(h_1,\,h_2,\,h_3)=(l_e,\,l_{\overline e}, 4l_x)$.

Figure~\ref{fig02} shows the resulting ternary luminosity diagram. An evenly-spaced grid of points
(at intervals $\delta l_e=\delta l_{\overline e}=\delta 4 l_x=0.1$) is superposed to guide the eye.
A red triangle marks the equipartition point $(l_e,\,l_{\overline e},\,4l_x)=(1/6,\,1/6,\,4/6)$. 
Departures from equipartition have often been considered \cite{Noneq}. 
For instance, Ref.~\cite{Luna} suggests to adopt factor-of-two-uncertainties
of the kind $1/2<l_e/l_x<2$ and $1/2<l_{\overline e}/l_x<2$; the corresponding ``allowed region''
is shown as a shaded inner diamond in Fig.~\ref{fig02}. In the same figure, the outer (light-shaded) 
diamond corresponds
to a more conservative allowed region, embracing factor-of-four uncertainties in both
$l_e/l_x$ and $l_{\overline e}/l_x$.

%%%%%%%%%%%%%%%%%%%%%%%%%%% FIGURE 3 %%%%%%%%%%%%%%%%%%%%%%%%%%%%%%%%
\newpage
\begin{figure}[t]
\centering
\vspace*{1mm}
\hspace*{23mm}
\epsfig{figure=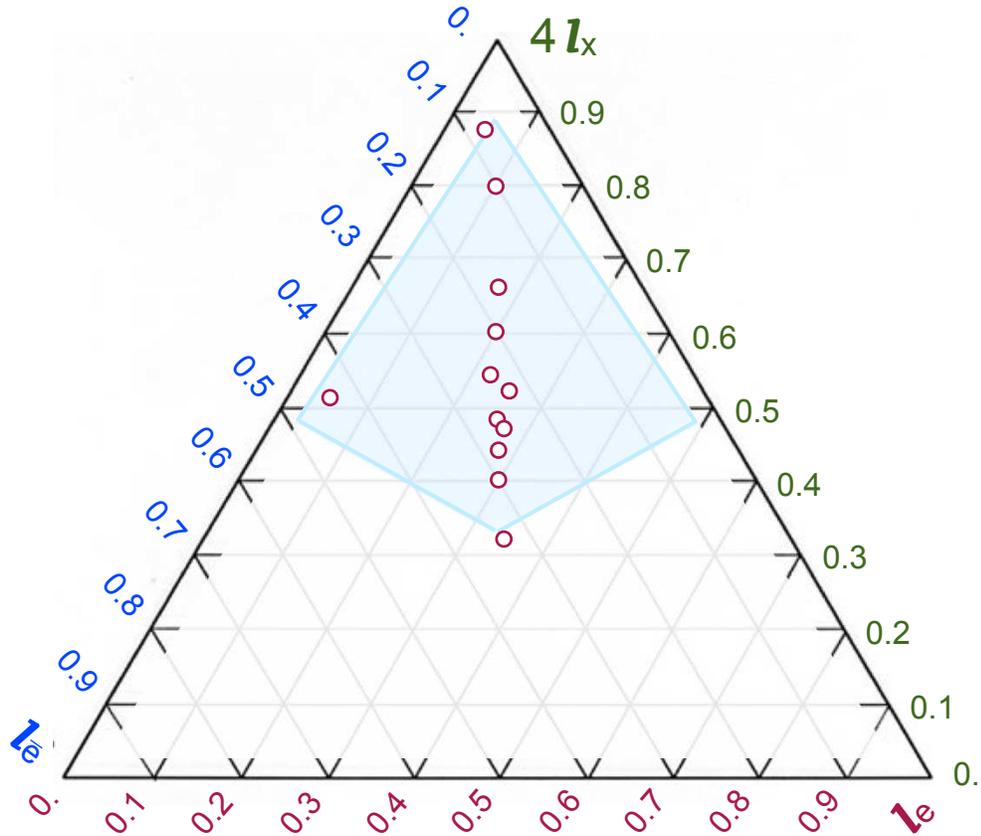,width=0.84\columnwidth}
\vspace*{-5mm}
\caption{Ternary luminosity diagram. Distribution of luminosities for a set of
supernova models, as taken from the compilation in Table~7.3 of Ref.~\protect\cite{Keil}.
\label{fig03}}
\end{figure}
%%%%%%%%%%%%%%%%%%%%%%%%%%%%%%%%%%%%%%%%%%%%%%%%%%%%%%%%%%%%%%%%%%%%%%

Figure~\ref{fig03} shows the distribution of fractional luminosities for a set
of various supernova simulation outputs, as taken from the compilation reported in Table~7.3 of Ref.~\cite{Keil}. Most
of the models cluster around the vertical line $l_{\overline e}\sim l_e$, except for one model
with very high ratio $l_{\overline e}/ l_e$ \cite{Hille}. The overall spread in $l_x$ is rather large. 
It should be noted that the above models refer to different post-bounce times and different
assumptions about SN (astro)physics; see \cite{Keil} for a critical discussion and comparison. In  any case,
all such models are basically included in the shaded diamond which corresponds to a factor-of-four 
uncertainty in $l_e/l_x$ and $l_{\overline e}/l_x$, and which will be referred to
as the ``phenomenologically interesting region.'' 

In this work we shall mainly 
focus on such restricted region for the detailed discussion of collective effects; however, 
for the sake of completeness, 
we have actually explored the full ternary diagram over all the inner points of the triangular grid,
for a total of 36 cases  (plus the equipartition one) in either hierarchies. 
The (computationally demanding) numerical investigation of all such different cases 
entails some assumptions and approximations, as described in the next Section.

\newpage

%%%%%%%%%%%%%%%%%%%%%%%%%%%%%%%%%%%%%%%%%%%%%%%%%%%%%%%%%%%%%%%%%%%%%%%%%%%%%%%%%%%%%%%%%%%%%%%%%%%%%%%%%
\section{Supernova Neutrino Framework}
\label{Sec3}
%%%%%%%%%%%%%%%%%%%%%%%%%%%%%%%%%%%%%%%%%%%%%%%%%%%%%%%%%%%%%%%%%%%%%%%%%%%%%%%%%%%%%%%%%%%%%%%%%%%%%

Observations and models of SN explosions suggest a diversity of both initial
and time-dependent features, and thus a relatively broad range
of possible inputs for SN neutrino  studies. We focus on specific SN inputs in terms of
luminosities, spectra, geometry, frequencies, evolution equations, and selected 
approximations, as described below.

\subsection{Neutrino luminosities and spectra}

In previous works \cite{Miri,Tamb,Three}, inspired by the SN model in \cite{Schir}, 
we have assumed a total SN binding energy $E_B\simeq 3\times 10^{53}$~erg, and a typical
luminosity decay timescale $\tau\simeq 3$~s. At early times ($t\ll\tau$), where large
deviations from the equipartition hypothesis may take place \cite{Noneq,Totani}, these
choices suggests a total luminosity 
%.....................................................................
\begin{equation}
L_\mathrm{tot} \simeq 10^{53}~\mathrm{erg/s}\ , 
\end{equation}
%...................................................................
which we fix as input, while
leaving free the fractional luminosities $l_\alpha=L_\alpha/L_\mathrm{tot}$.

Concerning average energies, we adopt  
typical early-time values \cite{Noneq,Totani},
%.....................................................................
\begin{equation}
\langle E_e \rangle = 10~\mathrm{MeV}, \ 
\langle E_{\overline e} \rangle = 12~\mathrm{MeV}, \ 
\langle E_x \rangle = 15~\mathrm{MeV}. \ 
\end{equation}
%...................................................................
Finally, the initial (normalized) energy spectra $\Phi_\alpha(E)$ are assumed to be thermal,
%.....................................................................
\begin{equation}
\Phi_\alpha(E) = \frac{2\beta_\alpha}{3\zeta_3}\,\frac{(\beta_\alpha E)^2}{e^{(\beta_\alpha E)}+1}\ ,
\end{equation}
%...................................................................
where $\zeta_3\simeq 1.202$ and $\beta_\alpha$ is an inverse temperature parameter $(\beta =1/T)$ \cite{Miri},
equal to
%.....................................................................
\begin{equation}
\beta_\alpha = \frac{c_{+}}{\langle E_\alpha \rangle}=\left\{
\begin{array}{ll}
0.315~\mathrm{MeV}^{-1} & (\alpha = e)\ ,\\
0.263~\mathrm{MeV}^{-1} & (\alpha = \overline e)\ ,\\
0.210~\mathrm{MeV}^{-1} & (\alpha = x)\ ,
\end{array}
\right. 
\end{equation}
%...................................................................
with
%.....................................................................
\begin{equation}
c_{+}=\frac{7\pi^4}{180\zeta_3}\simeq 3.151\ .
\end{equation}
%...................................................................

\subsection{Geometry}

We assume a bulb model \cite{Semi} with neutrinosphere radius
%..............
\begin{equation}
R_\nu=10~\mathrm{km}\ .
\end{equation}
%..................
We also adopt the so-called single-angle approximation, where
neutrino-neutrino interactions are averaged along
a single, radial trajectory \cite{Semi}. In this case, for $r>R_\nu$, the effective $\nu_\alpha$
number density per unit of volume and energy is given by \cite{Semi}
%.....................................................................
\begin{equation}
n_\alpha(r,\,E)=\frac{L_\mathrm{tot}l_\alpha}{4\pi R_\nu^2}\,\frac{\Phi_\alpha(E)}{\langle E_\alpha\rangle}\,g(r)\ ,
\end{equation}
%...................................................................
where $g(r)$ is a geometrical damping factor,
%.....................................................................
\begin{equation}
g(r)=\left[1-\sqrt{1-\left(\frac{R_\nu}{r}\right)^2}\,\right]^2\ ,
\end{equation}
%...................................................................
decreasing from unity to zero as $\sim\!1/r^4$. 
(See however \cite{Norm} for a somewhat different function $g(r)$ 
proposed in the single-angle limit.)
Integration over energy provides 
the effective $\nu_\alpha$ density per unit volume,
%.....................................................................
\begin{equation}
N_\alpha(r) = \int dE\,n_\alpha(r,\,E) = c(r)\,l_\alpha\, \beta_\alpha\ ,
\end{equation}
%...................................................................
where 
%.....................................................................
\begin{equation}
c(r) = \frac{L_\mathrm{tot}}{4\pi R^2_\nu}\,\frac{g(r)}{c_{+}}\ .
\end{equation}
%...................................................................

For definiteness, the absolute fluxes per unit of energy, 
%.....................................................................
\begin{equation}
F_{\alpha}(E) = \frac{L_\mathrm{tot}l_\alpha}{4\pi d^2}\frac{\Phi_\alpha(E)}{\langle E_\alpha\rangle}\ ,
\end{equation}
%...................................................................
will always be projected at a typical galactic-center distance of
%.....................................................................
\begin{equation}
d=10~\mathrm{kpc}\ ,
\end{equation}
%...................................................................
for both initial and intermediate spectra (at the end of collective effects).

\subsection{Frequencies}

In general, the most relevant frequencies in the context of SN neutrino collective
effects are the vacuum oscillation frequency $\omega$, the matter potential profile $\lambda(r)$, and 
the neutrino potential profile $\mu(r)$. We neglect the subdominant vacuum frequency
driven by the smallest (``solar'') $\delta m^2$ and the tiny $\nu_\mu$-$\nu_\tau$ 
matter potential difference which, as discussed in \cite{Three}, produce only very small effects
even at early times in our SN model; see also \cite{Esteban}.

The vacuum frequency reads
%.....................................................................
\begin{equation}
\omega = \frac{\Delta m^2}{2E} = \frac{5.07}{E/\mathrm{MeV}}\,[\mathrm{km}^{-1}]\ ,
\end{equation}
%...................................................................
where $\Delta m^2=2\times 10^{-3}$~eV$^2$ has been taken. We shall consider 
the energy range $E\in[0.1,\,80]$~MeV, corresponding to $\omega\in [0.06,\, 50]$~km$^{-1}$. 
Note that, as nicely shown in \cite{Mult}, the collective $\nu$ and $\overline\nu$ dynamics
merge in the limit $\omega\to 0$, namely at high $E$
(at least in an effective $2\nu$ description).

The matter potential $\lambda(r)=\sqrt{2}G_F N_e(r)$ (where $N_e$ is the electron
number density) can, to a large extent, be rotated away in the equations of motion \cite{Coll}. 
However, matter effects play a role, e.g., in 
delaying the start-up of collective effects \cite{Pend},  modifying some low-energy split features \cite{Tamb},  
generating later resonance effects for relatively large $\theta_{13}$ (see, e.g., \cite{Di08}), and possibly
inducing early decoherence \cite{Multidec}. 
By neglecting $\lambda(r)$, one trades such possible effects for a significant speed-up 
of the numerical calculations, which is a relevant gain in our systematic
exploration of the ternary diagram. 
Therefore, in the spirit of Ref.~\cite{Mult},  we choose to neglect $\lambda$ from the beginning, and to focus
on the dominant effects generated by the neutrino interaction potential $\mu$.

%%%%%%%%%%%%%%%%%%%%%%%%%%% FIGURE 4 %%%%%%%%%%%%%%%%%%%%%%%%%%%%%%%%
\newpage
\begin{figure}[t]
\centering
\vspace*{1mm}
\hspace*{23mm}
\epsfig{figure=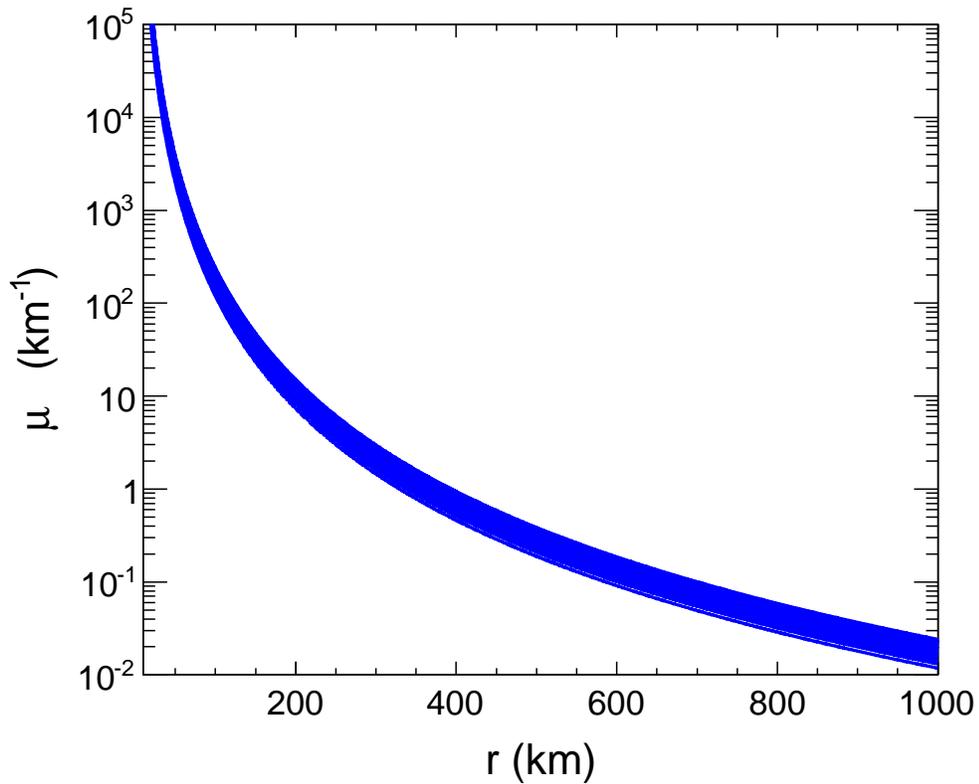,width=0.84\columnwidth}
\vspace*{-5mm}
\caption{Neutrino interaction potential $\mu(r)$ for our supernova model, in
the radial range $r\in[10,\,10^3]$~km. The vertical spread is a consequence of slight
variations of the curves $\mu(r)$ over different grid points of the ternary luminosity
diagram.
\label{fig04}}
\end{figure}
%%%%%%%%%%%%%%%%%%%%%%%%%%%%%%%%%%%%%%%%%%%%%%%%%%%%%%%%%%%%%%%%%%%%%%

The $\mu$ radial profile is given by
%................................................................
\begin{equation}
\mu(r)=\sqrt{2}G_F(N+\overline N)\ ,
\end{equation}
%.................................................................
where, as we shall argue below, the appropriate $\nu$ and  $\overline\nu$ number
densities ($N$ and $\overline N$, respectively)  refer to an 
effective $2\nu$ scenario, namely
%.........................................................................
\begin{equation}
N=N_e+N_x\ ,\ \overline N=N_{\overline e}+N_{\overline x}\ , 
\end{equation}
%........................................................................ 
so that 
%.....................................................................
\begin{equation}
\label{mu}
\mu(r)=c(r)\,(l_e\,\beta_e+l_{\overline e}\,\beta_{\overline e}+2\,l_x\,\beta_x)\ .
\end{equation}
%...................................................................

Figure~\ref{fig04} shows the corresponding family of $\mu(r)$ curves (too close to be distinguishable) 
in our adopted SN model, as obtained by varying the fractional luminosities $l_\alpha$ over the 36 grid points
in the ternary diagram.

Collective effects are expected to vanish when $\mu(r)$ is somewhat
below the smallest vacuum frequency considered, $\omega_\mathrm{min}\simeq 0.06$~km$^{-1}$.
We have verified that such vanishing is numerically realized within $r<10^{3}$~km
at any grid point; for safety, calculations are extended to $r=10^3$~km in all cases.

\newpage
\subsection{Flavor evolution formalism}

In the paper \cite{Three} we have performed detailed three-flavor calculations for our SN model
at the equipartition point for different times. The results of \cite{Three} show that, up to small 
effects at the percent (or lower) level, the collective 
dynamics basically involve only two neutrino species $(\nu_e,\,\nu_x)$, while the
$\nu_{\mu,\tau}$ combination orthogonal to $\nu_x$ remains a ``spectator'' neutrino species, 
and similarly for antineutrinos; see also \cite{Das3}. For instance, a typical spectral swap can be thought
to occur between $\nu_e$ and one of the two species $\nu_x$, the other being unaffected 
during collective evolution. The spectator species enters, however, in the 
later evolution at large radius via the usual $\theta_{12}$ mixing effects \cite{Das3}.
By focusing our attention on
the main collective effects within $r<10^{3}$~km, we can then 
reduce the full $3\nu$ evolution to an effective $2\nu$ one, namely: $3\nu=(\nu_e,\,\nu_x)\oplus \nu_x$,
where the relevant $2\nu$ subspace ($\nu_e,\,\nu_x$) is governed by $(\pm \Delta m^2,\,\theta_{13})$. As in
\cite{Three}, we choose a default value
%..........
\begin{equation}
\sin^2\theta_{13}=10^{-6}\ .
\end{equation}
%..........
We have explicitly verified, for several (non-equipartition) test cases in our SN scenario,
that the effective $2\nu$ calculations [with $\mu(r)$ defined as in Eq.~(\ref{mu})]
reproduce the results obtained from a full $3\nu$ approach \cite{Three}
within percent accuracy.

Concerning the $2\nu$ framework, we adopt the same notation
of \cite{Miri} in terms of single-mode Bloch vectors $\mathbf{P}$ and $\overline\mathbf{P}$
in flavor space, 
appropriately extended to generic fractional luminosities
$l_\alpha$. In particular, the initial conditions for the global polarization 
vectors $\mathbf{J}$, $\overline \mathbf{J}$ and for their difference $\mathbf{D}=\mathbf{J}-
\overline \mathbf{J}$, are given by:
%...............................................................................
\begin{eqnarray}
\mathbf{J} &=& \frac{1}{N+\overline N}\int dE\,(n_e-n_x)\,\mathbf{z} = 
\frac{l_e\beta_e-l_x\beta_x}{l_e\beta_e+l_{\overline e}\beta_{\overline e}+2l_x\beta_x}\,\mathbf{z} \ ,\\
\overline\mathbf{J} &=& \frac{1}{N+\overline N}\int dE\,(n_{\overline e}-n_{\overline x})\, \mathbf{z} = 
\frac{l_{\overline e}\beta_{\overline e}-l_x\beta_x}{l_e\beta_e+l_{\overline e}\beta_{\overline e}+2l_x\beta_x}\,\mathbf{z} \ ,\\
\mathbf{D} &=& \frac{1}{N+\overline N}\int dE\,(n_e-n_{\overline e})\,\mathbf{z} = 
\frac{l_e\beta_e-l_{\overline e}\beta_{\overline e}}{l_e\beta_e+l_{\overline e}\beta_{\overline e}+2l_x\beta_x}\,\mathbf{z} \ .
\end{eqnarray}
%...............................................................................
Note that the integrand in $\mathbf{J}$ is positive (negative) for energies below (above) the
crossing energy where the $\nu_e$ and $\nu_x$ fluxes are equal, and similarly
for $\overline \mathbf{J}$; see also Fig.~\ref{fig01}. The $D_z$ component of $\mathbf{D}$ is
a constant of motion, corresponding to $\nu_e-\overline\nu_e$ number conservation \cite{Pend}.

Finally, we remind that a potential energy ${\cal V}$ can be attached to the system,
%.....................................................................
\begin{equation}
{\cal V} \simeq \mp \mathbf{z}\,(\mathbf{W}+\overline\mathbf{W})\ ,
\end{equation}
%...................................................................
where the upper (lower) sign refers to normal (inverted) hierarchy. The global vectors
$\mathbf{W}$ and $\overline\mathbf{W}$ are defined by energy integrals analogous to 
$\mathbf{J}$ and $\overline\mathbf{J}$ in the above equations, but 
with a further factor $\omega$ in the integrand \cite{Miri}. Therefore, also such vectors receive
positive (negative) contributions below (above) the crossing energies. The role of
the crossing energies in understanding single and multiple splits has been emphasized in \cite{Mult}.

%%%%%%%%%%%%%%%%%%%%%%%%%%% FIGURE 5 %%%%%%%%%%%%%%%%%%%%%%%%%%%%%%%%
\newpage
\begin{figure}[t]
\centering
\vspace*{1mm}
\hspace*{23mm}
\epsfig{figure=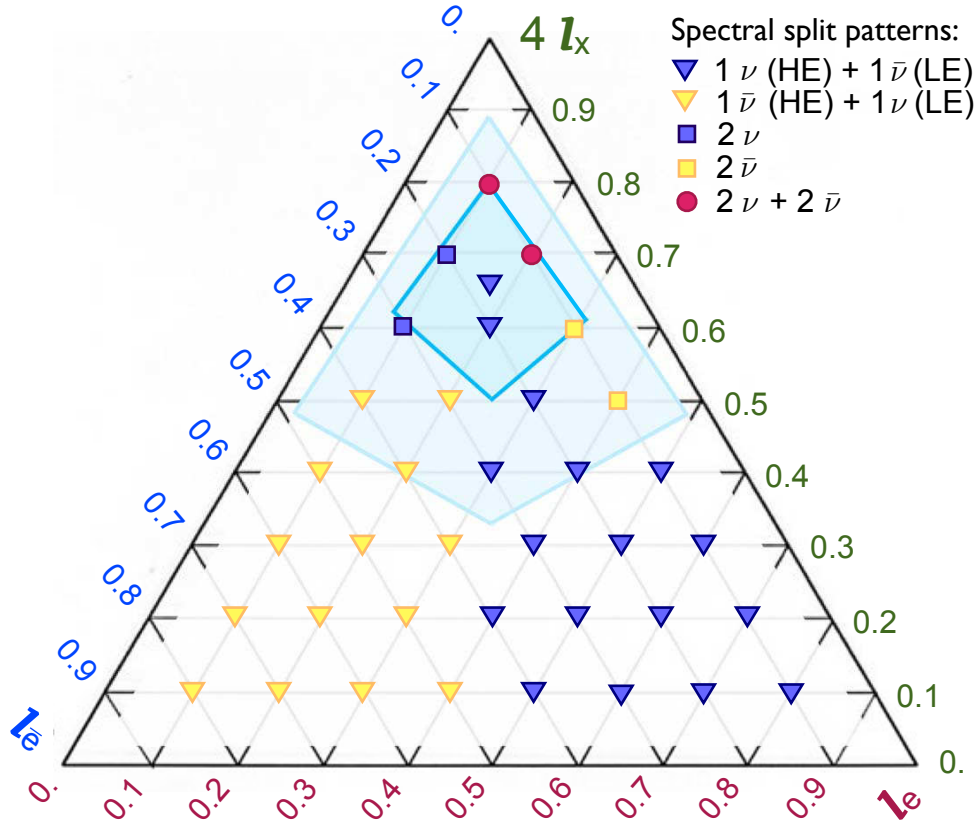,width=0.84\columnwidth}
\vspace*{-5mm}
\caption{Spectral split patterns in inverted hierarchy. Blue triangles:
one ``high-energy'' $\nu$ split and one ``low-energy'' $\overline\nu$ split.
Blue squares: two $\nu$ splits (and no $\overline \nu$ split). Yellow triangles
and squares: as before, but with $\nu$ and $\overline\nu$ interchanged.
Red circles: two splits for both $\nu$ and  $\overline\nu$.
\label{fig05}}
\end{figure}
%%%%%%%%%%%%%%%%%%%%%%%%%%%%%%%%%%%%%%%%%%%%%%%%%%%%%%%%%%%%%%%%%%%%%%

%%%%%%%%%%%%%%%%%%%%%%%%%%%%%%%%%%%%%%%%%%%%%%%%%%%%%%%%%%%%%%%%%%%%%%%%%%%%%%%%%%%%%%%%%%%%%%%%%%%%%
\section{Results for Inverted Hierarchy: Description}
\label{Sec4}
%%%%%%%%%%%%%%%%%%%%%%%%%%%%%%%%%%%%%%%%%%%%%%%%%%%%%%%%%%%%%%%%%%%%%%%%%%%%%%%%%%%%%%%%%%%%%%%%%%%%%

Figure~\ref{fig05} shows the qualitative spectral split patterns emerging from
our numerical exploration of the ternary luminosity diagram,
in the case of inverted hierarchy. In the lower half of the diagram, corresponding
to relatively low $\nu_x$ luminosity ($4l_x<0.5$), we always find one $\nu$ and 
one $\overline\nu$ split. More precisely, the blue triangles on the right correspond to
one dominant $\nu$ split at high energy (HE) plus a minor $\overline\nu$ split at
low energy (LE), qualitatively similar to Fig.~\ref{fig01}; for the yellow triangles
on the left, the situation is reversed for $\nu$ and $\overline\nu$. As the $\nu_x$
luminosity increases, some $1\nu~(\mathrm{HE})+1\nu~(\mathrm{LE})$ split cases
survive (blue triangles), including the equipartition point.  However, these
cases are now flanked, on the left, by a couple of points where a
double $\nu$ split occurs (blue squares) and, on the right, by two points
with a double $\overline\nu$ split (yellow squares) plus two points with a double
split for both $\nu$ and $\overline\nu$ (red circles). Therefore, a complex
phenomenology of multiple splits \cite{Mult} emerges in the ternary diagram, even if restricted 
to the region of phenomenological interest (the largest shaded diamond). In particular,
small departures from luminosity equipartition can abruptly change
the split patterns.

%%%%%%%%%%%%%%%%%%%%%%%%%%% FIGURE 6 %%%%%%%%%%%%%%%%%%%%%%%%%%%%%%%%
\newpage
\begin{figure}[t]
\centering
\vspace*{1mm}
\hspace*{23mm}
\epsfig{figure=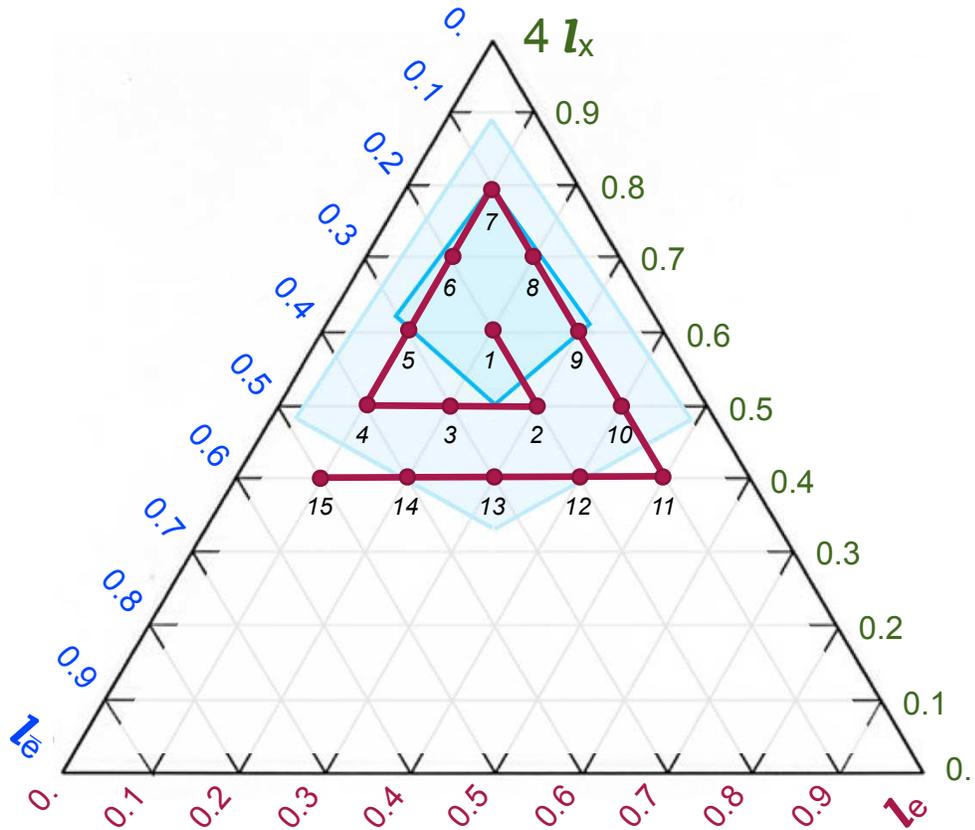,width=0.84\columnwidth}
\vspace*{-5mm}
\caption{Representative grid points (numbered from 1 to 15), which cover
all the observed split patterns. The points define a trajectory which
starts close to the equipartition case and then spirals outwards.
\label{fig06}}
\end{figure}
%%%%%%%%%%%%%%%%%%%%%%%%%%%%%%%%%%%%%%%%%%%%%%%%%%%%%%%%%%%%%%%%%%%%%%

In order to better appreciate these changes, quantitative results 
will be shown for representative grid points. Figure~\ref{fig06} shows
fifteen selected points, forming
a trajectory which starts close to the equipartition case and then
spirals outwards. This trajectory allows to explore the region of phenomenological interest,
and to cover all the five (qualitatively different) single and double split patterns described 
in the comment to Fig.~\ref{fig05}. The corresponding spectra are discussed next.

\begin{figure}[t]
\centering
\vspace*{1mm}
\hspace*{23mm}
\epsfig{figure=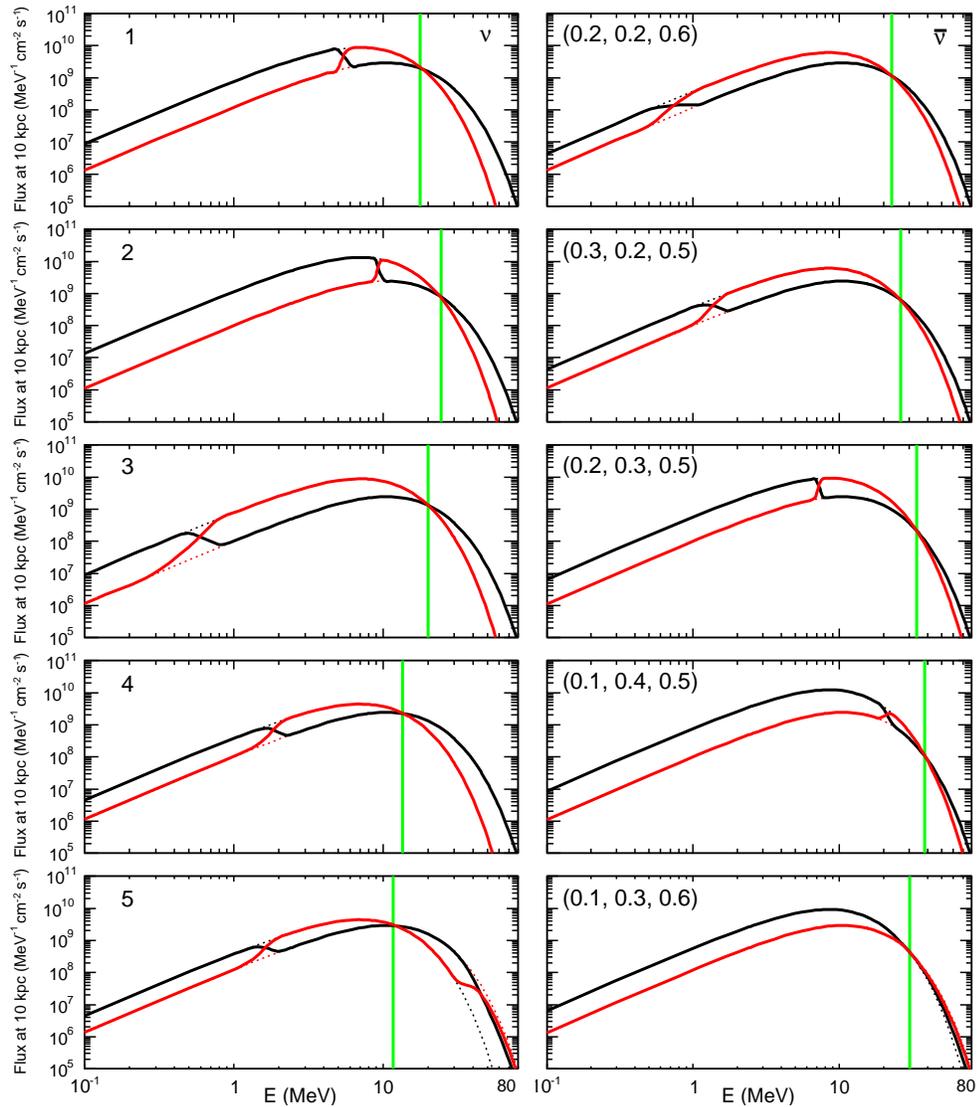,width=0.84\columnwidth}
\vspace*{-5mm}
\caption{Neutrino (left) and antineutrino (right) energy spectra at the end of 
collective effect, for the points 1--5 in Fig.~\protect\ref{fig06}. The corresponding fractional luminosities
$(l_e,\,l_{\overline e},\,l_x)$ are indicated in each right panel. Vertical green lines
mark the crossing energies where the $\nu_e$ (or $\overline\nu_e$) and $\nu_x$ fluxes 
are equal.
\label{fig07}}
\end{figure}

\begin{figure}[t]
\centering
\vspace*{1mm}
\hspace*{23mm}
\epsfig{figure=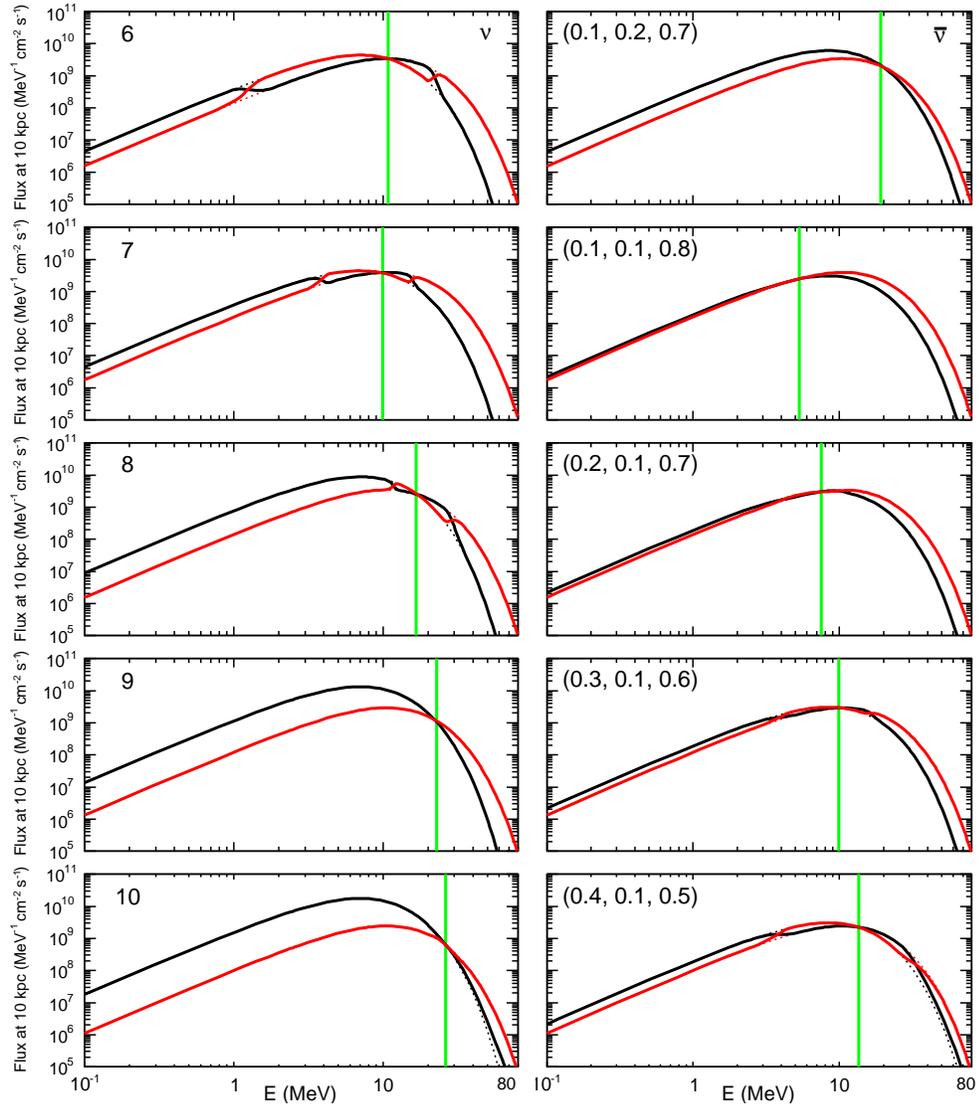,width=0.84\columnwidth}
\vspace*{-5mm}
\caption{As in Fig.~\protect\ref{fig07}, but for points 6--10. 
\label{fig08}}
\end{figure}

\begin{figure}[t]
\centering
\vspace*{1mm}
\hspace*{23mm}
\epsfig{figure=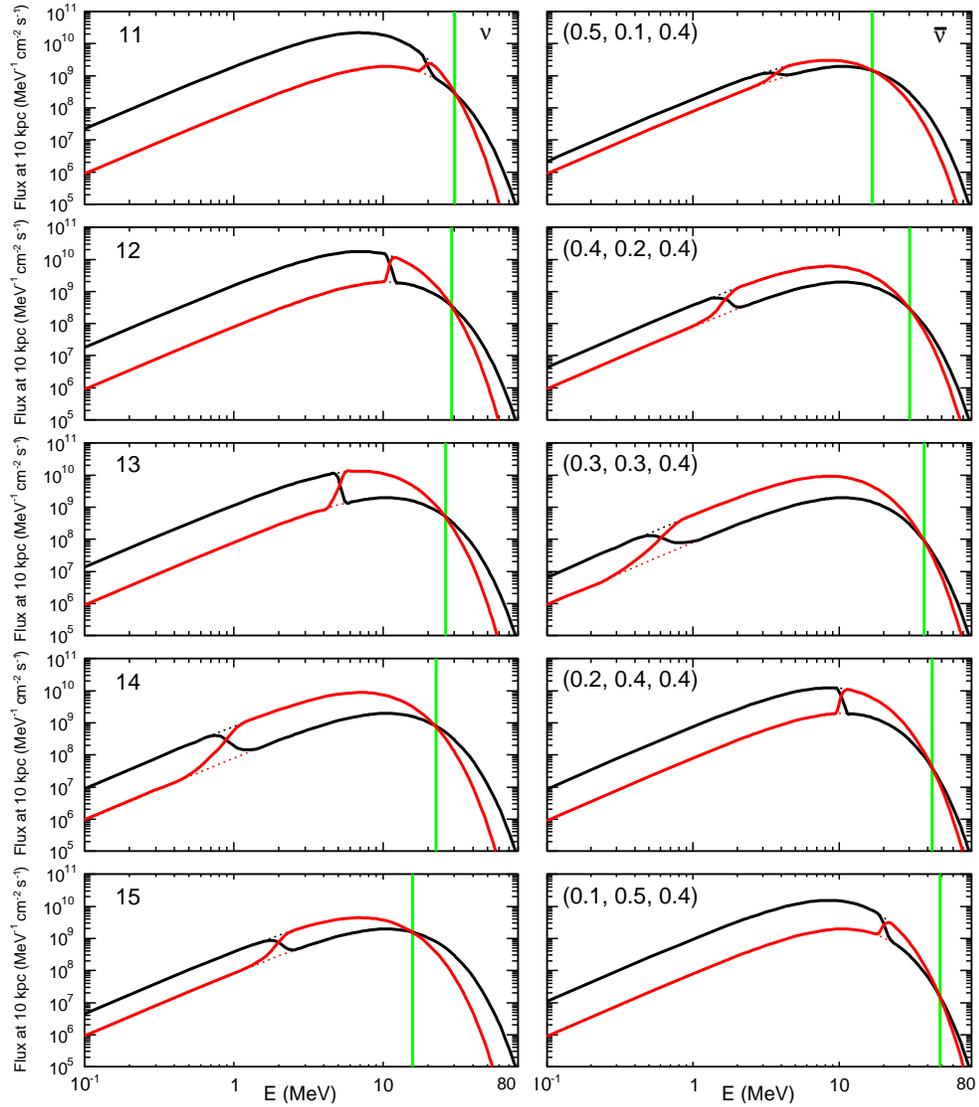,width=0.84\columnwidth}
\vspace*{-5mm}
\caption{As in Fig.~\protect\ref{fig07}, but for points 11--15. 
\label{fig09}}
\end{figure}

Figures~\ref{fig07}, \ref{fig08} and \ref{fig09} show the oscillated fluxes
of neutrino (left) and antineutrinos (right) per unit energy, at the 
end of collective effects ($r<10^3$~km). The black and red solid curves
refer to $e$-flavor and $x$-flavor, respectively; dotted curves represent
unoscillated fluxes. Vertical green lines
mark crossing energies, where fluxes of different flavor intersect,
and flavor changes---if any---are unobservable. It is worth noticing that, in all cases,
the unoscillated fluxes have qualitatively similar features, despite the
large differences in relative luminosities: the $\nu_x$ flux is always
lower (higher) than the $\nu_e$ flux, for energies below (above) the crossing energy $E_c$,
which ranges between 10 and 30~MeV; and similarly for antineutrinos, with $\overline E_c$
in the range 5--30~MeV. Nevertheless, oscillated fluxes show 
large differences from point to point.
In all panels, we indicate
the point number (from 1 to 15) on the left, and the luminosity values
$(l_e,\,l_{\overline e},\,4l_x)$ on the right.

%%%%%%%%%%%%%%%%%%%%%%%%%%% FIGURE 7 %%%%%%%%%%%%%%%%%%%%%%%%%%%%%%%%
\newpage
%%%%%%%%%%%%%%%%%%%%%%%%%%%%%%%%%%%%%%%%%%%%%%%%%%%%%%%%%%%%%%%%%%%%%%

In Fig.~\ref{fig07}, both the 1st and the 2nd point show a split pattern very similar
to the equipartition case in Fig.~\ref{fig01}, with a sharp HE $\nu$ split and a broader,
LE $\overline\nu$ split. The situation is suddenly reversed 
in the 3rd and 4th point, showing  a sharp HE $\overline\nu$ split and a broader,
LE $\nu$ split. The split pattern changes again in the 5th point, where 
a second split appears for $\nu$'s above the crossing energy, while no evident
flavor change emerges for $\overline\nu$'s (up to small effects in the upper tail of the spectrum).

%%%%%%%%%%%%%%%%%%%%%%%%%%% FIGURE 8 %%%%%%%%%%%%%%%%%%%%%%%%%%%%%%%%
\newpage
%%%%%%%%%%%%%%%%%%%%%%%%%%%%%%%%%%%%%%%%%%%%%%%%%%%%%%%%%%%%%%%%%%%%%%

In Fig.~\ref{fig08}, the 6th point shows a split pattern similar to the last
one in the previous figure, namely, a double $\nu$ split and no $\overline\nu$ split.
However, in both the 7th and 8th point, a double split occurs also for $\overline\nu$'s,
although the closeness of the $\overline\nu_e$ and $\overline\nu_x$ spectra 
make it difficult to appreciate it graphically (it will be shown in a different way below). 
The 9th and 10th point show a more evident $\overline\nu$ double split, but no
$\nu$ flavor change (up to small upper-tail effects in the last case).
The grid points 5--10 exhaust the cases where we find a double split in 
$\nu$ and/or $\overline\nu$ spectra.

%%%%%%%%%%%%%%%%%%%%%%%%%%% FIGURE 9 %%%%%%%%%%%%%%%%%%%%%%%%%%%%%%%%
\newpage
%%%%%%%%%%%%%%%%%%%%%%%%%%%%%%%%%%%%%%%%%%%%%%%%%%%%%%%%%%%%%%%%%%%%%%

In Fig.~\ref{fig09}, the cases from 11 to 13 show a
pattern with a sharp HE $\nu$ split and a broader
LE $\overline\nu$ split, while the situation is suddenly reversed in
the last two (14th and 15th) cases. It should be noted that the 
sudden exchange of roles between $\nu$ and $\overline\nu$ 
across cases 13 and 14 occurs despite the fact the unoscillated
spectra and the crossing energies are very similar. 

In general,
the results of Figs.~\ref{fig07}--\ref{fig09} suggest the
existence of a sort of ``phase transitions'' in the ternary diagram, with
abrupt changes in the split patterns when 
the initial luminosities cross some ``phase boundaries,''
which we shall try to identify in the next Section.

%%%%%%%%%%%%%%%%%%%%%%%%%%% FIGURE 10 %%%%%%%%%%%%%%%%%%%%%%%%%%%%%%%%
\newpage
\begin{figure}[t]
\centering
\vspace*{1mm}
\hspace*{23mm}
\epsfig{figure=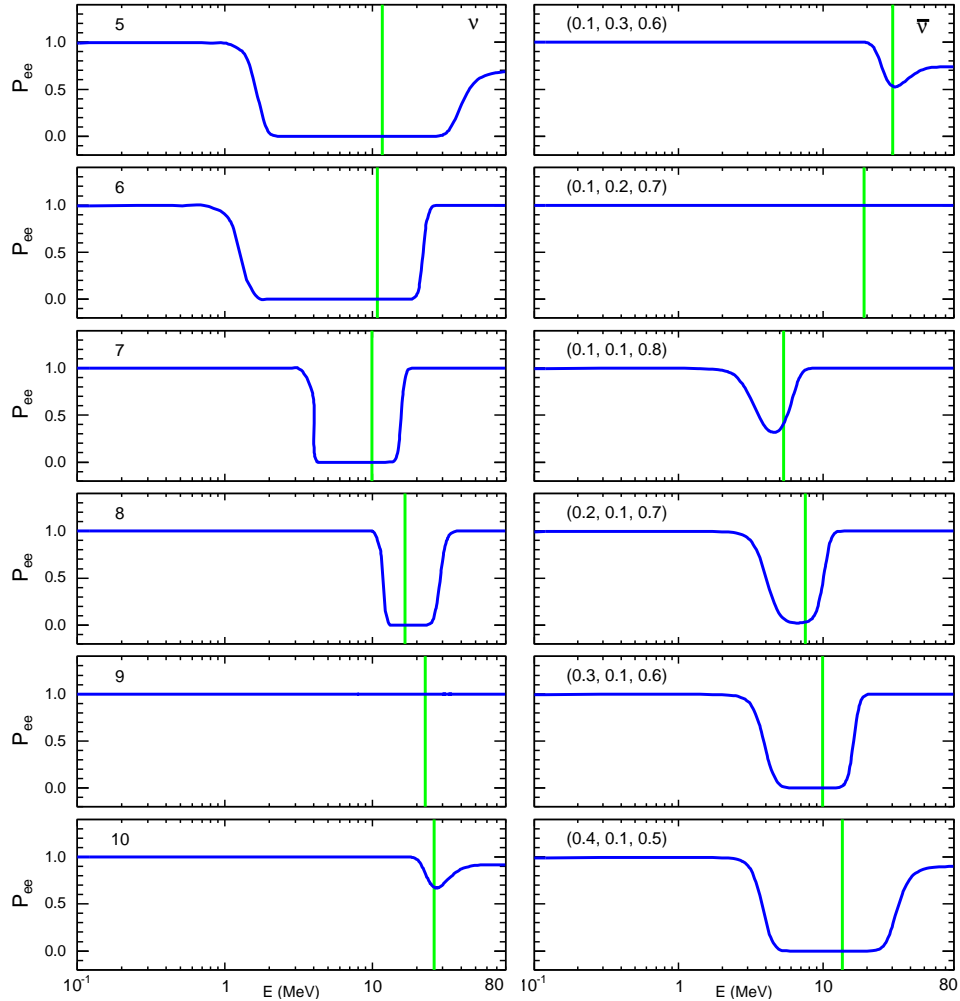,width=0.82\columnwidth}
\vspace*{-5mm}
\caption{Survival probability $P_{ee}$ for neutrinos (left) and
antineutrinos (right) in each of six points 5--10 of Fig.~\protect\ref{fig06}, 
where a double split is observed for neutrinos and/or antineutrinos.
\label{fig10}}
\end{figure}
%%%%%%%%%%%%%%%%%%%%%%%%%%%%%%%%%%%%%%%%%%%%%%%%%%%%%%%%%%%%%%%%%%%%%%

Finally, Fig.~\ref{fig10} shows the $e$-flavor survival probability profile
for neutrinos (left) and antineutrinos (right), for the six grid points numbered
from 5 to 10 (see Fig.~\ref{fig06}), where double splits occur in either $\nu$
or $\overline\nu$ or both (see Fig.~\ref{fig05}). All double splits flank the
crossing energy (vertical green line), in agreement with the arguments of \cite{Mult}.
As also noted in  \cite{Mult}, the values of $P_{ee}$ for $\nu$ and $\overline\nu$ coincide at high $E$ (i.e., for $\omega \to 0$).

A double-double split is evident
only in cases 7 and 8. In cases 5 and 10, instead, there seems to be a partly developed 
``would-be double split'' for $\overline\nu$ and $\nu$, respectively. In the spirit
of \cite{Mult}, it is tempting
to attribute the suppression of such emerging splits to the relatively high
(thus unfavorable \cite{Mult}) values of the associated crossing energies; however,
this argument is insufficient to explain the total absence of $\overline\nu$ and $\nu$
double splits in cases 6 and 9, where the associated crossing energies
are favorably lower. In the next Sec.~\ref{Sec5}, we shall develop different
arguments, trying to relate (at least some of)
such spectral features to global initial conditions and potential energy minimization.
Additional comments will be provided in Sec.~\ref{Sec7}.

%%%%%%%%%%%%%%%%%%%%%%%%%%% FIGURE 11 %%%%%%%%%%%%%%%%%%%%%%%%%%%%%%%%
\newpage
\begin{figure}[t]
\centering
\vspace*{1mm}
\hspace*{23mm}
\epsfig{figure=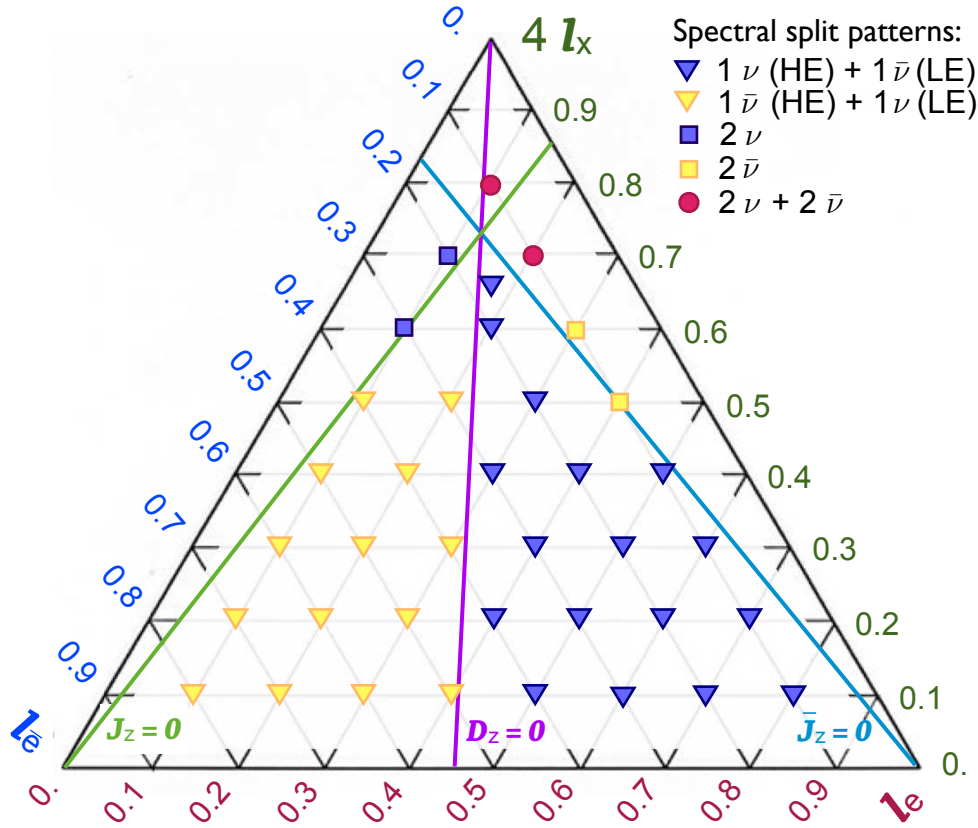,width=0.84\columnwidth}
\vspace*{-5mm}
\caption{Spectral split patterns, superposed to the three lines characterized
by $J_z=0$ (green, with $J_z>0$ below the line), by $\overline J_z=0$
 (blue, with $\overline J_z>0$ below the line), and where $D_z=0$ (violet, with 
 $D_z>0$ on the right of the line). 
 The lines mark transitions to different split patterns.
\label{fig11}}
\end{figure}
%%%%%%%%%%%%%%%%%%%%%%%%%%%%%%%%%%%%%%%%%%%%%%%%%%%%%%%%%%%%%%%%%%%%%%

%%%%%%%%%%%%%%%%%%%%%%%%%%%%%%%%%%%%%%%%%%%%%%%%%%%%%%%%%%%%%%%%%%%%%%%%%%%%%%%%%%%%%%%%%%%%%%%%%%%%%
\section{Results for Inverted Hierarchy: Interpretation}
\label{Sec5}
%%%%%%%%%%%%%%%%%%%%%%%%%%%%%%%%%%%%%%%%%%%%%%%%%%%%%%%%%%%%%%%%%%%%%%%%%%%%%%%%%%%%%%%%%%%%%%%%%%%%%

The authors of \cite{Mult} propose a very interesting attempt to explain analytically the development
and the stabilization of spectral splits in terms of adiabatic invariants, energy minimization,
and localization of the crossing energies (with details to appear in a forthcoming paper). 
Here we aim at providing another 
viewpoint on the interpretation of single and double spectral splits, with emphasis on the initial
orientation of the global polarization vectors and on some non-adiabatic features.

Initially, the global polarization vectors $\mathbf{J}$, $\overline\mathbf{J}$ and $\mathbf{D}$
are oriented along the $\mathbf{z}$ axis. Figure~\ref{fig11} displays the curves (straight
lines in our SN model) where $J_z=0$ (green), $\overline J_z=0$ (blue) and $D_z=0$ 
(violet) at the neutrinosphere, superposed to the same marked grid of points of Fig.~\ref{fig05}.
 We surmise that sign changes of $J_z$, $\overline J_z$ and
$D_z$ across these lines mark ``phase transitions'' across different split patterns
in the ternary diagram, as argued below for the various cases reproduced
from Fig.~\ref{fig05} (blue and yellow triangles and squares, red circles). 

\subsection{Cases with $1\nu (\mathit{HE})+1\overline\nu(\mathit{LE})$ split: blue triangles}

These cases, which include the equipartition one, are characterized by  $J_z>0$, $\overline J_z>0$, and $D_z>0$. 
In other words, both $\mathbf{J}$ and the smaller vector $\overline \mathbf{J}$ are 
initially aligned upwards, in the unstable pendulum position \cite{Pend}. 
As already argued in \cite{Miri}, the flavor pendulum dynamics
would bring both vectors downwards, but is impeded by conservation of $D_z$; the
simplest way to minimize the potential energy is then to invert 
the vector $\overline\mathbf{ J}$ and an equivalent fraction
of the  vector $\mathbf{J}$, leading to a sharply defined ``high-energy'' neutrino split
(whose development can be described by detailed adiabatic constructions \cite{Smirnov}).

However, the ``low-energy'' antineutrino split seem to have a different, nonadiabatic 
origin, as already argued in \cite{Tamb} and suggested by its broader features. 
In the present work (where the 
effects  of the matter potential $\lambda$ are ignored), we find that the $\overline\nu$
modes below the split energy $\overline E_s$ basically experience a usual ``resonant transition'' 
when $-\omega+\mu D_z \simeq 0$, and then decouple from the collective dynamics. In the
energy range considered ($E\geq 0.1$~MeV), this resonance is highly nonadiabatic, and leads
to a two-level hopping probability $P_C\simeq 1$, corresponding to $P_{ee}\simeq 1$. [We have explicitly
verified the expected decrease of $P_C$ and $P_{ee}$ at lower energies, as well as their
dependence on the $d\log \mu/d r$ profile.] In other words, the low-energy neutrino modes
resonate with the interaction potential, and do not participate in later collective
swaps. We find that this phenomenon roughly takes place during (and a little bit after)
the so-called period of synchronized oscillations \cite{Sync} (which
is then followed by ``bipolar'' oscillations \cite{Analysis}). Defining $\mu_{s}$ as the value of 
$\mu$ at the transition from synchronized to bipolar oscillations \cite{Miri}, the 
 antineutrino energy modes which remain pinned upwards are roughly determined by
the condition $\omega >\mu_s D_z$. In our numerical experiments, such condition identifies the antineutrino
split energies within a factor of about 2--3; 
in particular, we find that the highest values of such energies are reached for the
grid points closest to the $D_z=0$ line.

\subsection{Cases with $1\nu (\mathit{LE})+1\overline\nu(\mathit{HE})$ split: yellow triangles}

These cases are similar to the previous one, but the roles of $\nu$ and $\overline \nu$ are exchanged,
since $D_z<0$ and thus ${\overline J}_z$ is larger than $J_z$. The 
dominant HE split, dictated by $D_z$-conservation, occurs for antineutrinos. The
inversion of low-energy neutrino modes appears to be impeded by early
nonadiabatic resonances for $\omega>-\mu_s D_z$.

\subsection{Cases with $2\nu$ splits: blue squares}

In these two cases (identified as n.~5 and 6 in Figs.~\ref{fig06} and \ref{fig10}), 
$\mathbf{J}$ is oriented upwards, in the unstable position,
while $\overline \mathbf{J}$ is oriented downwards and is thus already stable.
However, the vector $\mathbf{J}$ cannot be inverted, even partially (i.e., with
a single split),  due to $D_z$ conservation: the two vectors are ``locked'' and
remain basically equal in modulus and orientation at the end of collective
effects, as we have explicitly verified. However, one can still invert 
a fraction of the neutrino spectra ``symmetrically'' with respect to $E_c$ \cite{Mult},
so as to preserve the modulus of $\mathbf{J}$ while decreasing the value of
$W_z$ (and thus the potential energy). The double neutrino split appears
thus a highly constrained way to lower the energy of the system, when
the global polarization vectors of $\nu$ and $\overline\nu$ are 
basically locked. 

Finally, we note that the point identified as n.~5 in Figs.~\ref{fig06} and \ref{fig10} 
represents a ``borderline'' case, being almost on top of the line at $J_z=0$. This
specific feature might be at the origin of the partial ``double split,''
barely emerging in the antineutrino spectrum for this case (see Fig.~\ref{fig10}).

\subsection{Cases with $2\overline\nu$ splits: yellow squares}

These two cases (numbered as points n.~9 and 10 in Figs.~\ref{fig06} and \ref{fig10}) 
are analogous to the previous ones (n.~6 and 5, respectively), but the
role of $\nu$ and $\overline\nu$ is exchanged. So, the same considerations apply, including
the ``borderline'' features of point 10 which sits on top of the $\overline J_z=0$ line.

A third case (numbered as point n.~8 in Figs.~\ref{fig06} and \ref{fig10}) should also fall
in the same category, being characterized by the same initial signs of the global
polarization vectors. However, for this point (the lowest red circle in Fig.~\ref{fig11}) 
we find double splits for both $\nu$ and $\overline\nu$ (see Fig.~\ref{fig10}). We have 
been unable to  explain this ``exception'' by sharpening energy minimization arguments.
See, however, Sec.~\ref{Sec7} for further comments.

\subsection{Cases with $2\nu+2\overline\nu$ splits: red circles}

These cases include the mentioned point n.~8 (which escapes a simple explanation)
and point n.~7 (the highest red circle in Fig.~\ref{fig11}), which can be instead more easily interpreted. 
The latter point is the only one, in our grid,  having both $\mathbf{J}$ and $\overline\mathbf{J}$
initially oriented downward, in the stable pendulum position. Then, the only way to further
minimize the potential energy (without altering $D_z$) is to generate double splits
which preserve the moduli of $\mathbf{J}$ and $\overline\mathbf{J}$ while
decreasing the values of $W_z$ and $\overline W_z$, which is indeed what happens
at the end of the collective evolution (see Fig.~\ref{fig10}).

\subsection{Summary for inverted hierarchy}

In inverted hierarchy, both single and double splits may appear
in either the $\nu$ or the $\overline\nu$ spectra, depending on the initial distribution
of the total luminosity over the species ($\nu_e,\, \overline\nu_e,\, \nu_x$). The split
patterns observed numerically admit (with one
exception) a relatively simple interpretation in terms of initial conditions 
for the global polarization vectors $\mathbf{J}$,
$\overline\mathbf{J}$ and $\mathbf{D}$, supplemented by arguments related
to the flavor pendulum dynamics and its potential energy minimization. 
Transitions to different split patterns emerge when crossing the lines 
$J_z=0$, $\overline J_z=0$ and $D_z=0$ in the ternary luminosity diagram.

%%%%%%%%%%%%%%%%%%%%%%%%%%% FIGURE 12 %%%%%%%%%%%%%%%%%%%%%%%%%%%%%%%%
\newpage
\begin{figure}[t]
\centering
\vspace*{1mm}
\hspace*{23mm}
\epsfig{figure=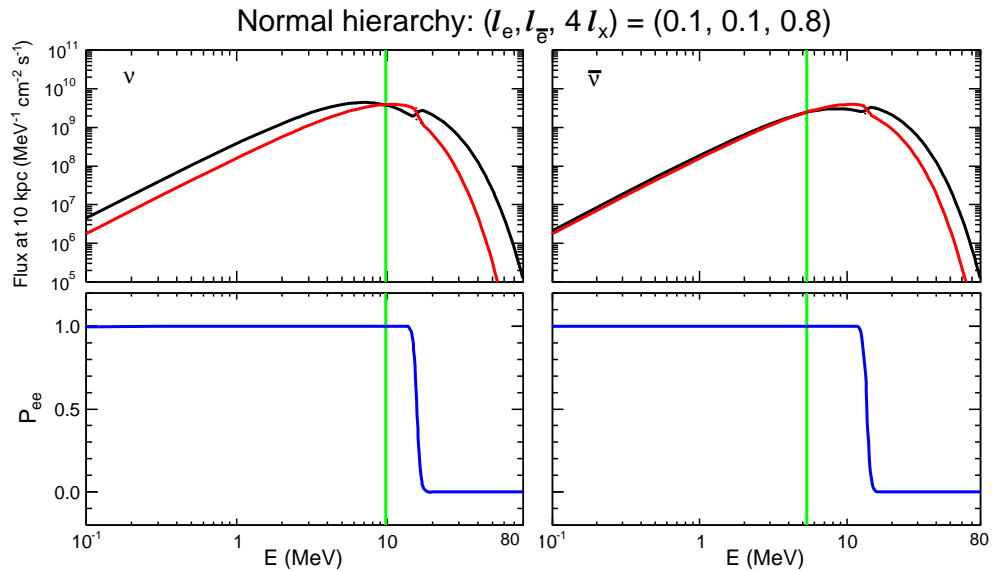,width=0.84\columnwidth}
\vspace*{-5mm}
\caption{Normal hierarchy: spectra for the single grid point where splits
are observed. For all other  grid points, collective effects do not change
the original spectra.
\label{fig12}}
\end{figure}
%%%%%%%%%%%%%%%%%%%%%%%%%%%%%%%%%%%%%%%%%%%%%%%%%%%%%%%%%%%%%%%%%%%%%%

%%%%%%%%%%%%%%%%%%%%%%%%%%%%%%%%%%%%%%%%%%%%%%%%%%%%%%%%%%%%%%%%%%%%%%%%%%%%%%%%%%%%%%%%%%%%%%%%%%%%%
\section{Results for Normal Hierarchy}
\label{Sec6}
%%%%%%%%%%%%%%%%%%%%%%%%%%%%%%%%%%%%%%%%%%%%%%%%%%%%%%%%%%%%%%%%%%%%%%%%%%%%%%%%%%%%%%%%%%%%%%%%%%%%%

In normal hierarchy, the flavor pendulum dynamics is generally trivial,
 since the initial conditions are typically very close to
the stable equilibrium position \cite{Pend}. However, an interesting counterexample
has been provided in \cite{Mult} for a situation with relatively large $\nu_x$ luminosity.

In our ternary  diagram, we find that collective effects produce
no flavor change in normal hierarchy, except for a single grid point---the one with the
highest $\nu_x$ luminosity. In this peculiar case, 
both $\mathbf{J}$ and $\overline\mathbf{J}$ start in the
unstable  position (now corresponding to downward orientation), and then
partially reverse, in order to minimize energy
while preserving $D_z$. As a result, single splits emerge in both $\nu$ and $\overline\nu$ 
spectra, as shown in Fig.~\ref{fig12}, in qualitative agreement with the normal-hierarchy case discussed in \cite{Mult}.

%%%%%%%%%%%%%%%%%%%%%%%%%%%%%%%%%%%%%%%%%%%%%%%%%%%%%%%%%%%%%%%%%%%%%%%%%%%%%%%%%%%%%%%%%%%%%%%%%%%
\section{Comments on a ``more adiabatic'' scenario}
\label{Sec7}
%%%%%%%%%%%%%%%%%%%%%%%%%%%%%%%%%%%%%%%%%%%%%%%%%%%%%%%%%%%%%%%%%%%%%%%%%%%%%%%%%%%%%%%%%%%%%%%%%%%%

In Sec.~\ref{Sec5}, the proposed
interpretation of the double split patterns emerging in Fig.~\ref{fig10} 
shows that initially stable (downward-aligned) global polarization vectors 
may be (cases 7, 8) or may not be (cases 5, 6, and 9, 10) further stabilized by a double split. In view of the
results in \cite{Mult}, it is worthwhile to
investigate if the absence of some ``would-be double splits'' can be related to
incomplete adiabaticity. To this purpose, we examine a hypothetical
scenario, where the function $\mu(r)$ is radially stretched by a factor of ten 
above the neutrinosphere,
%...............
\begin{equation}
\label{adiab}
\mu(r)\to \mu\left(R_\nu+\frac{r-R_\nu}{10}\right)\ .
\end{equation}
%................

%%%%%%%%%%%%%%%%%%%%%%%%%%% FIGURE 13 %%%%%%%%%%%%%%%%%%%%%%%%%%%%%%%%
\newpage
\begin{figure}[t]
\centering
\vspace*{1mm}
\hspace*{23mm}
\epsfig{figure=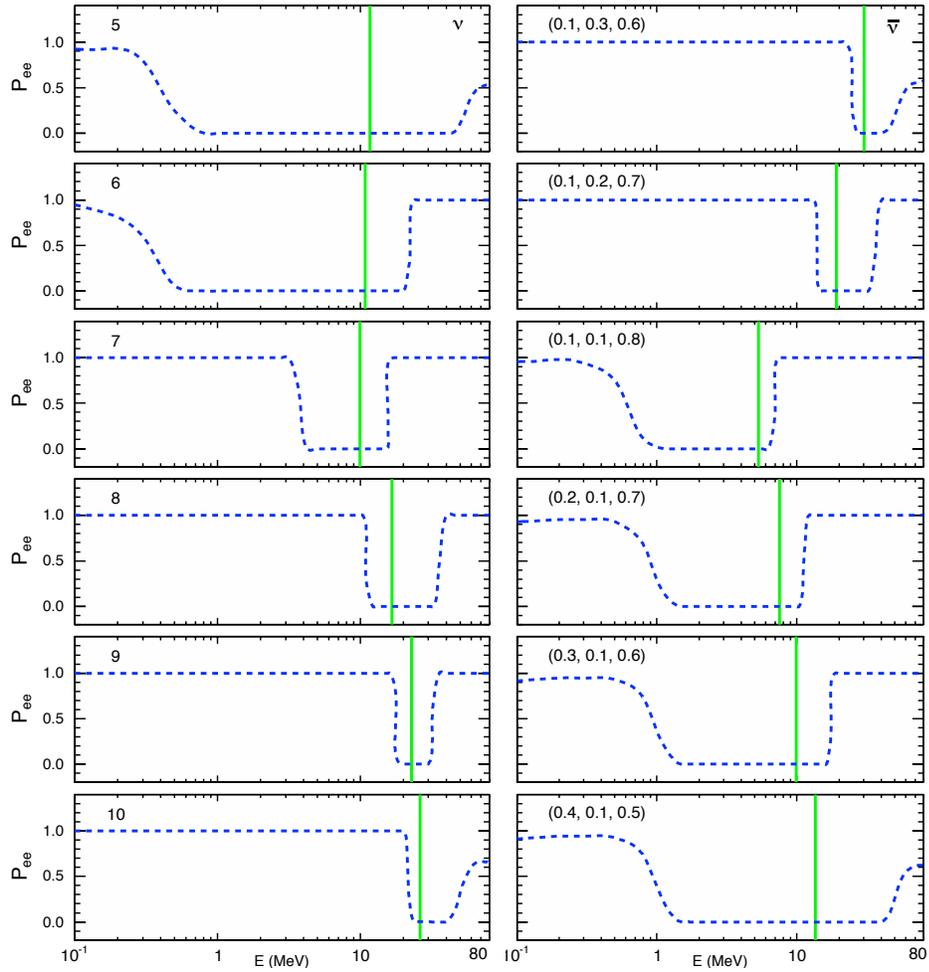,width=0.81\columnwidth}
\vspace*{-1mm}
\caption{As in Fig.~\protect\ref{fig10}, but for the ``more adiabatic'' $\mu(r)$ profile
in Eq.~(\protect\ref{adiab}).
\label{fig13}}
\end{figure}
%%%%%%%%%%%%%%%%%%%%%%%%%%%%%%%%%%%%%%%%%%%%%%%%%%%%%%%%%%%%%%%%%%%%%%
The main results for such ``more adiabatic'' scenario are summarized in Fig.~\ref{fig13}.
A comparison with the analogous Fig.~\ref{fig10} shows that  the
increased adiabaticity make $2\nu+2\overline\nu$ splits emerge in all cases 5--10. In particular, for cases 5 and 6, we find
that the ``new'' double splits provide a further (although quantitatively small)
decrease of the  potential energy for $\overline\nu$'s, which already start from stable
initial conditions ($\overline J_z<0$); and analogously for $\nu$'s in cases 9 and 10. 
In other words, the $\nu$ or $\overline\nu$ systems seem to have ``enough  time'' to relax further,
and to minimize an already low potential energy. Note also that the split energies are somewhat 
different in Fig.~\ref{fig13} as compared with Fig.~\ref{fig10}, especially on the low-energy side,
where we have already argued (Sec.~5.1) that nonadiabatic resonances play a prominent role.  
In general, more adiabatic profiles for $\mu(r)$ [such as in Eq.~(\ref{adiab})]
seem to cancel qualitative differences
among the different double split cases commented in Secs.~5.3, 5.4 and 5.5, which  merge
in the single $2\nu+2\overline\nu$ case described in Sec.~5.5. Conversely,
all single-split cases remain qualitatively similar, as we have explicitly verified by numerical exploration
of the ternary diagram.

%%%%%%%%%%%%%%%%%%%%%%%%%%% FIGURE 14 %%%%%%%%%%%%%%%%%%%%%%%%%%%%%%%%
\newpage
\begin{figure}[t]
\centering
\vspace*{1mm}
\hspace*{23mm}
\epsfig{figure=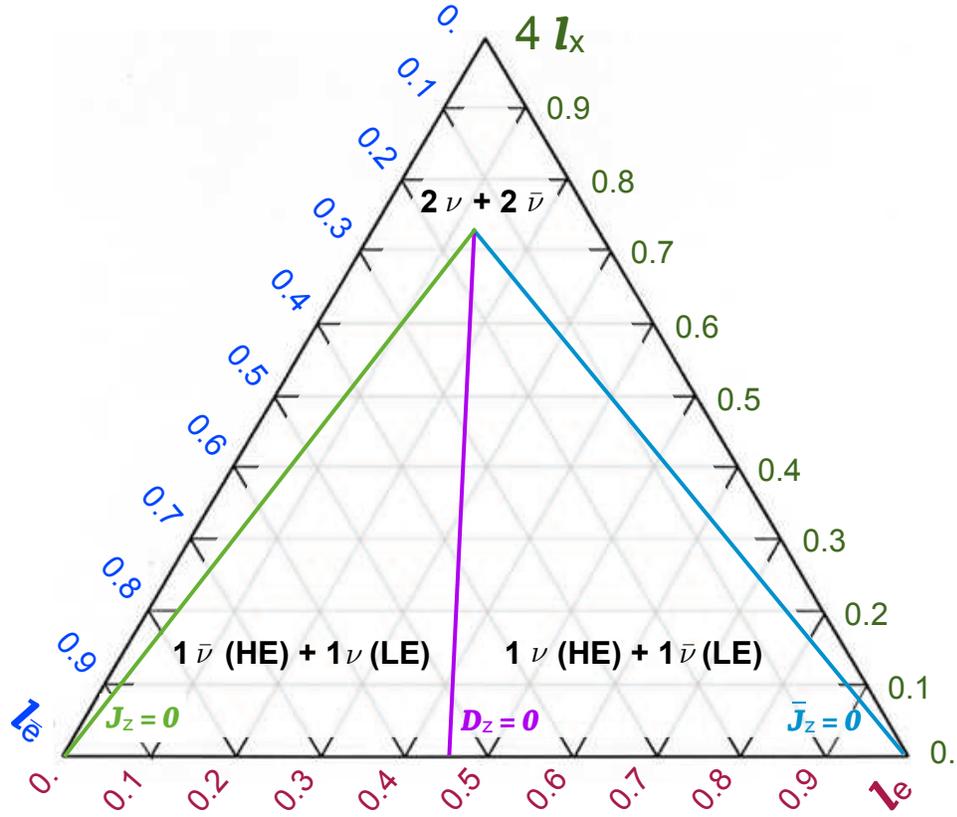,width=0.84\columnwidth}
\vspace*{-0mm}
\caption{Spectral split patterns for a scenario with increased
adiabaticity (see the text for details). The lines, as in Fig.~\protect\ref{fig11},  mark transitions to different split patterns.
\label{fig14}}
\end{figure}
%%%%%%%%%%%%%%%%%%%%%%%%%%%%%%%%%%%%%%%%%%%%%%%%%%%%%%%%%%%%%%%%%%%%%%

Figure~\ref{fig14} shows the resulting, ``adiabatic'' spectral split patterns, to be compared
with the ones in Fig.~\ref{fig11}. It appears that there are at least three distinct ``phases'' in the
ternary diagram,
with either single splits (at low/high or high/low energy) or double splits in the $\nu$ and $\overline\nu$
spectra. Within the $2\nu+2\overline\nu$ region of Fig.~\ref{fig14}, possible sub-phases seem
to emerge as a result of incomplete adiabaticity in Fig.~\ref{fig11}. 

The equipartition case is close to the crossing point of the three curves in
Figs.~\ref{fig11} and \ref{fig14}. Therefore, in the inverted hierarchy case,
relatively small variations in the fractional
luminosities, or in the (non)adiabatic character of the neutrino evolution, may induce
significant variations in the observable spectral split patterns. 

Concerning the normal hierarchy case, we do not find qualitatively new features by
using the modified function $\mu(r)$ in Eq.~(\ref{adiab}), as compared with the discussion in Sec.~\ref{Sec6}.

\newpage

%%%%%%%%%%%%%%%%%%%%%%%%%%%%%%%%%%%%%%%%%%%%%%%%%%%%%%%%%%%%%%%%%%%%%%%%%%%%%%%%%%%%%%%%%%%%%%%%%%%%%
\section{Summary and Prospects}
\label{Sec8}
%%%%%%%%%%%%%%%%%%%%%%%%%%%%%%%%%%%%%%%%%%%%%%%%%%%%%%%%%%%%%%%%%%%%%%%%%%%%%%%%%%%%%%%%%%%%%%%%%%%%%

In core-collapse
supernovae, collective effects involving high-density (anti)neutrinos
continues to reveal surprising features. In particular, relatively small variations of the luminosities
associated to $\nu_e$, $\overline\nu_e$ and $\nu_x$ appear to trigger 
abrupt changes from single- to double-split features in the energy spectra.

We have investigated 
the effect of generic variations of the fractional luminosities
$(l_e,\, l_{\overline e}, \, 4l_x)$ with respect to the usual ``energy equipartition'' case 
 $(1/6,\,1/6,\,4/6)$, within an early-time supernova scenario with fixed thermal 
spectra and total luminosity. The constraint
$l_e+l_{\overline e}+4l_x=1$ has been embedded in a ``ternary luminosity diagram,'' 
which has been numerically explored 
over an evenly-spaced grid of points.

In inverted hierarchy, we have found both single- and double-split cases for
either neutrino or antineutrino spectra, and have
proposed an interpretation of these patterns in terms of 
initial orientations of the global flavor polarization vectors $\mathbf{J}$ and $\overline\mathbf{J}$,
and of minimization of the potential energy, constrained by lepton number
conservation ($D_z=J_z-{\overline J}_z=\mathrm{const}$). 
In particular, the curves defined by $J_z=0$, ${\overline J}_z=0$ and $D_z=0$ appear
to provide ``phase transition boundaries'' between regions with single- and double-split 
features. 
The regions where at least one double split occurs (for either 
$\nu$ or $\overline\nu$) can actually merge into a single one with $2\nu+2\overline\nu$
splits, if adiabaticity is increased by design. It turns out that
the luminosity equipartition point is relatively close to the crossing point of the
three separation curves; therefore, relatively small variations of the fractional luminosities
can produce qualitatively different swaps and quantitatively significant changes
in the final spectra, at the end of the collective favor
evolution.

On one hand, these results may provide a handle to reconstruct the original 
luminosities $(l_e,\, l_{\overline e}, \, 4l_x)$, if the associated spectral split 
patterns can be observed in future galactic SN explosions. On the other hand, they
complicate the calculation of the diffuse supernova neutrino background, since past 
SN events may well have different relative luminosities and thus
different spectral split features.
Even within a single core-collapse SN event, there might be transitions
from single- to double-split spectra (or viceversa) during the first few seconds after bounce,
as a results of time-dependent changes in the relative luminosities, average energies,
(non)adiabatic features,
and possible departures from thermal spectra. 

The analysis of this rich phenomenology is just
at the beginning, and calls for a deeper theoretical interpretations and for more refined
numerical explorations, in order to get the most from observable supernova neutrino events.
Concerning the results presented in this work, we plan to refine them by
focusing on the phenomenologically interesting region close to the equipartition
point, where further 
insights on the spectral split patterns may be expected, in both normal and inverted hierarchy, 
by adopting a more general SN framework and a denser sampling grid.

%%%%%%%%%%%%%%%%%%%%%%%%%%%%%%%%%%%%%%%%%%%%%%%%%%%%%%%%%%%%%%%%%%%%%%%%%%%%%%%%%
\section*{Acknowledgments}
 %%%%%%%%%%%%%%%%%%%%%%%%%%%%%%%%%%%%%%%%%%%%%%%%%%%%%%%%%%%%%%%%%%%%%%%%%%%%%%%%%%%
This work is supported in part by
the Italian  ``Istituto Nazionale di Fisica Nucleare'' (INFN) and  ``Ministero dell'Istruzione, 
dell'Universit\`a e 
della Ricerca''  (MIUR) through the ``Astroparticle Physics'' research project. We also acknowledge
support by the E.U.\ (ENTApP network). I.T.\ thanks the organizers of the TAUP'09 Conference,
where preliminary results of this
work were presented, for kind hospitality.

%%%%%%%%%%%%%%%%%%%%%%%%%%%%%%%%%%%%%%%%%%%%%%%%%%%%%%%%%%%%%%%%%%%%%%
\section*{References} %%%%%%%%%%%%%%%%%%%%%%%%%%%%%%%%%%%%%%%%%%%%%%%%
%%%%%%%%%%%%%%%%%%%%%%%%%%%%%%%%%%%%%%%%%%%%%%%%%%%%%%%%%%%%%%%%%%%%%%

\end{document}